\documentclass[aps,prl,superscriptaddress,showpacs,floatfix,twocolumn,byrevtex]{revtex4-1}
\pdfoutput=1
\usepackage{amstext}
\usepackage{amsfonts}
\usepackage{amssymb}
\usepackage{bm}
\usepackage{color}
\usepackage{dcolumn}
\usepackage{graphicx}
\usepackage{hyperref}

\begin{document}
%
% Title
%
\title{Phononic gaps in the charged incommensurate planes of \boldmath
  Sr$_{14}$Cu$_{24}$O$_{41}$ \unboldmath}
%
% Authors and affiliations
%
\author{V.~K.~Thorsm{\o}lle}
\email{verner.thorsmolle@epfl.ch}
\affiliation{Laboratory for Photonics and Interfaces, \'Ecole Polytechnique F\'ed\'erale de
  Lausanne (EPFL), CH-1015, Switzerland}
\affiliation{D\'{e}partement de Physique de la Mati\`{e}re Condens\'{e}e,
  Universit\'{e} de Gen\`{e}ve, CH-1211 Gen\`{e}ve 4, Switzerland}
\author{C. C. Homes}
\email{homes@bnl.gov}
\author{A. Gozar}
\email{agozar@bnl.gov}
\affiliation{Condensed Matter Physics and Materials Science Department,
  Brookhaven National Laboratory, Upton, New York 11973, USA}
\author{G. Blumberg}
\affiliation{Department of Physics and Astronomy, Rutgers,
  The State University of New Jersey, Piscataway, New Jersey 08854, USA}
\author{J. L. M. van Mechelen}
\author{A. B. Kuzmenko}
\affiliation{D\'{e}partement de Physique de la Mati\`{e}re Condens\'{e}e,
  Universit\'{e} de Gen\`{e}ve, CH-1211 Gen\`{e}ve 4, Switzerland}
\author{S. Vanishri}
\author{C. Marin}
\affiliation{CEA Grenoble, INAC, SPSMS, IMAPEC, 17 rue des Martyrs, 38054 Grenoble, France}
\author{H. M. R{\o}nnow}
\affiliation{Laboratory for Quantum Magnetism, \'Ecole Polytechnique F\'ed\'erale de
  Lausanne (EPFL), CH-1015, Switzerland}
\date{\today}
%
% abstract
%
\begin{abstract}
The terahertz (THz) excitations in the quantum spin-ladder system Sr$_{14}$Cu$_{24}$O$_{41}$
have been determined along the {\em c}-axis using THz time-domain, Raman and infrared
spectroscopy.  Low-frequency infrared and Raman active modes are observed above and below
the charge-ordering temperature $T_{\rm co} \simeq 200$~K over a narrow interval $\simeq 1 - 2$~meV
($\simeq 8 - 16$~cm$^{-1}$). A new infrared mode at $\simeq 1$~meV
develops below $\simeq 100$~K.  The temperature dependence of these modes shows that
they are coupled to the charge- and spin-density-wave correlations in this system.
These low-energy features are conjectured to originate in the gapped sliding-motion of
the chain and ladder sub-systems, which are both incommensurate and charged.
\end{abstract}

% 71.45.Lr 	Charge-density-wave systems (see also 75.30.Fv Spin-density waves)
% 72.80.Le 	Polymers; organic compounds (including organic semiconductors)
% 78.30.-j 	Infrared and Raman spectra
% 78.70.Gq 	Microwave and radio-frequency interactions
%

\maketitle

\begin{figure}[t]
\centering
\includegraphics[width=0.9\columnwidth]{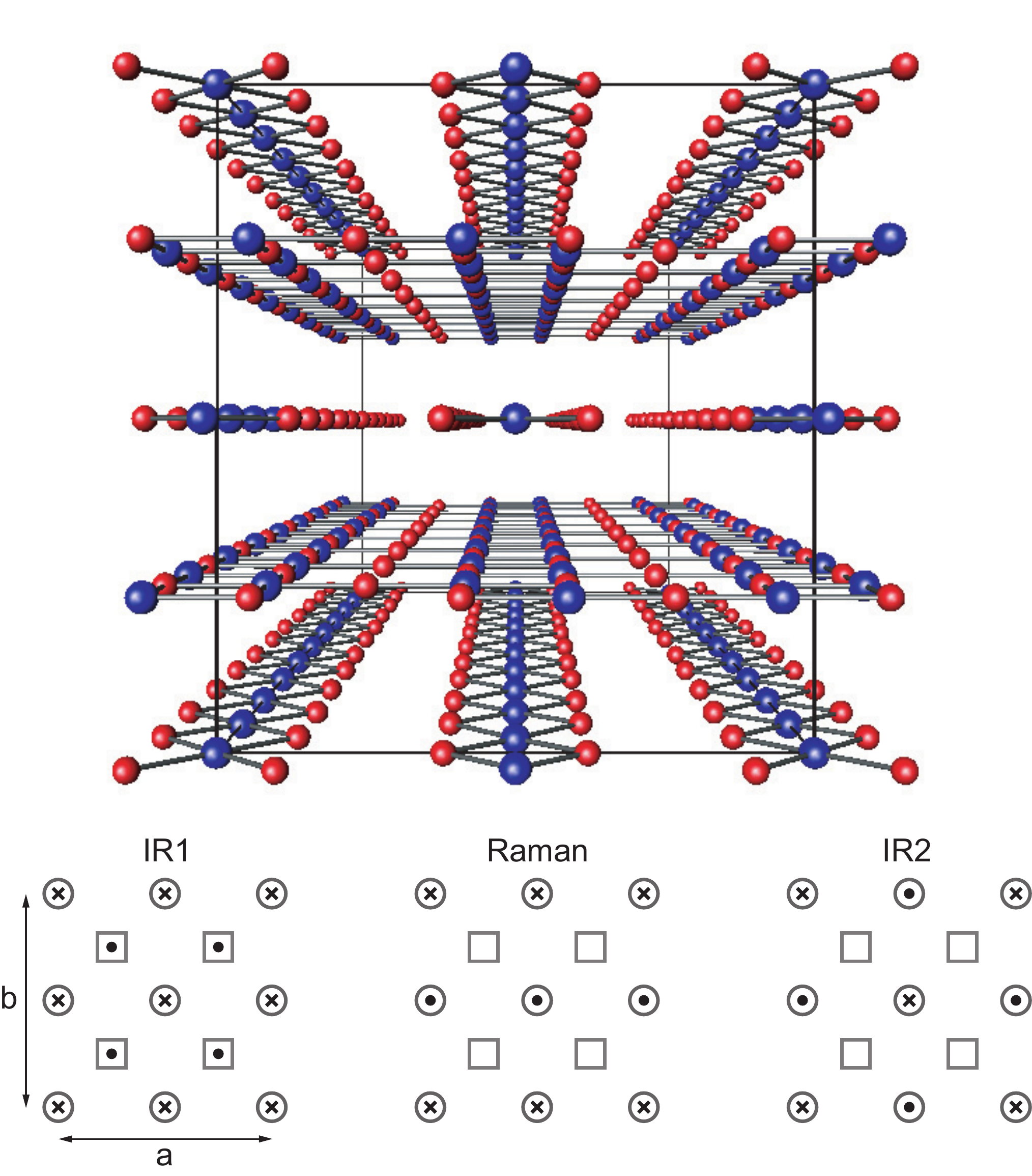}
\caption{Perspective view of the ideal unit cell of Sr$_{14}$Cu$_{24}$O$_{41}$ (the
Sr atoms are omitted) in the {\em a-b} plane viewed along the {\em c} axis showing the
stacking of the CuO$_2$ chains and the Cu$_2$O$_{3}$ ladders.  Beneath the unit
cell are the symbolic representations of the low-frequency infrared and Raman modes
in the ladders and chains (see text for details); the chains are depicted by the circles
($q_{\rm ch}\simeq -1.4\,e$), and the ladders by the squares ($q_{\rm ld}\simeq 2\,e$).
The dots and crosses refer to the oscillatory motion in and out of the {\em a-b}
plane, respectively.}%
\label{fig:modes}%
\end{figure}

%
% Introduction
%
More than three decades ago, new normal modes were predicted to occur in ionic materials with
incommensurate (IC) layers which can slide past each other \cite{Theodorou1978,Theodorou1980,
Axe1982,Finger1983}. These new degrees of freedom allow separate phonons in each subsystem at
high frequencies with a crossover to slow oscillations due to relative sliding motions of the two
almost rigid subsystems at ultra-low frequencies, leading effectively to an extra acoustic mode. If the IC layers are charged these sliding modes become gapped due to the restoring Coulomb forces. These modes are the ionic complements of the electronic plasmons in metals and their dynamics also  resemble the sliding motion in density wave (DW) systems \cite{Gruner1994}. Thus far,unambiguous experimental evidence for sliding gapped acoustic mode resonances has remained, to our knowledge, elusive.
A promising avenue of investigation is the low-dimensional quantum spin-ladder system
Sr$_{14}$Cu$_{24}$O$_{41}$ containing such substructures in the form of Cu$_2$O$_3$
ladders and one-dimensional (1D) CuO$_2$ chains \cite{McCarron1988,*Siegrist1988}.
The chains and ladders run parallel along the {\em c}-axis with the rungs of the ladders along
the {\em a}-axis \cite{McCarron1988,*Siegrist1988}, shown in Fig.~\ref{fig:modes}. The
two subsystems are structurally IC, resulting in a buckling along the
{\em c}-axis with a period $c=27.5\,{\rm \AA}\,\simeq 10\,c_{\rm ch} \simeq 7\,c_{\rm ld}$,
where $c_{\rm ch}$ and $c_{\rm ld}$ represent the lattice constants for the chain and
ladder subcells, respectively.

This intrinsically hole-doped material exhibits a variety of unusual charge, magnetic
and vibrational phenomena \cite{Vuletic2006} that have been probed by several
techniques, among them magnetic resonance \cite{Takigawa1998}, neutron scattering
\cite{Matsuda1999} and resonant x-ray scattering \cite{Abbamonte2004,Rusydi2006,Rusydi2008}.
The unusual DW order is attributed to cooperative phenomena driven and stabilized
by charge and spin correlations, in conjunction with the IC lattice
degrees of freedom.
Low-energy features spanning frequencies from the kHz to the THz range and
associated with the DW dynamics have also been observed.  However, while the
microwave  data have been consistently interpreted in terms of screened DW
relaxational dynamics \cite{Blumberg2002,Gorshunov2002,Gozar2003,Vuletic2003},
the nature of the excitations in the THz regime, which are seen to be strongly
coupled to the charge/spin ordering, remains controversial \cite{Blumberg2002,Gorshunov2002,
Gozar2003,Vuletic2003,Abbamonte2004,Vuletic2005,Choi2006,Rusydi2006,Rusydi2007,Rusydi2008}.
An important and open question is, what are the salient spectroscopic
features in the $\simeq 1$~meV energy region?

%
% What is done in this work...
%
In this Letter we demonstrate using three experimental techniques that low-energy
collective modes in Sr$_{14}$Cu$_{24}$O$_{41}$ consist of one Raman and one infrared
(IR) active excitation present in the $5-300$~K temperature region, as well as a new
mode appearing well below the charge-ordering temperature (T$_{\rm co}\simeq 200$~K)
whose temperature-dependent behavior is tracked by terahertz time-domain spectroscopy
(THz-TDS) \cite{Thorsmolle2007}. An intuitive interpretation of these modes in relation
to ``phononic gaps'' opened by Coulomb interactions in IC lattices and
coupled to the DW ordering is able to consistently explain the range and relative
energies of these excitations which are illustrated in  Fig.~\ref{fig:modes}.

%
% Figure 1
%

%
% Experimental details
%
Single crystals of Sr$_{14}$Cu$_{24}$O$_{41}$ were grown using the traveling-solvent
floating-zone method.  The sample was oriented and cut to $4\,{\rm mm} \times 6\,{\rm mm}$,
with a thickness of 440 $\mu$m and the {\em a}- and {\em c}-axes in the plane.
%
% At room temperature it is generally accepted to be $n_{\rm ch}\approx 5$ in the chains
% and $n_{\rm ld}\approx 1$ in the ladders \cite{Nucker2000}.
%
The THz-TDS experiments (TPI spectra 1000, TeraView Ltd.) were performed in transmission
geometry with the sample mounted inside an optical cryostat capable of reaching 5~K. To
obtain the complex conductivity and the transmittance window of the time-domain signals
lasting beyond 60~ps (containing several Fabry-Perot internal sample reflections), a
complete Drude-Lorentz time-domain analysis study is presented, in contrast to simple
frequency-inversion \cite{Brauer2009}.
The polarized reflectance was measured over a wide frequency range using an {\em in-situ}
evaporation technique \cite{Homes1993}.  The complex conductivity is determined from a
Kramers-Kronig analysis of the reflectance \cite{Dresselbook} requiring extrapolations
in the $\omega\rightarrow 0$ limit; above 250~K a Hagen-Rubens form is employed
$R(\omega) \propto 1-\sqrt{\omega}$, while below this temperature the reflectance
is assumed to be constant, $R(\omega \rightarrow 0) \simeq 0.68 - 0.76$.
The Raman measurements were performed in ({\em cc}) and ({\em aa}) polarizations as
described in Ref.~\onlinecite{Gozar2003}.  The direction of propagation of the light
was perpendicular to the {\em a-c} plane for all measurements.
%
%The complex optical response has been modeled using the Drude-Lorentz form of the
%complex dielectric function
%
%\begin{equation}
%   \tilde{\varepsilon}(\omega)=\varepsilon_{\infty}-\frac{\omega^2_{p,D}}{\omega^2+i\gamma_D\omega} +
%   \sum_{j} \frac{\omega^2_{p,j}}{\omega^2_{j}-\omega^2-i\gamma_j\omega}.
%\end{equation}
%
%Here $\varepsilon_{\infty}$ is the high-frequency dielectric constant, $\omega_{p,D}$
%and $\gamma_D$ the plasma frequency and width of the Drude component, while $\omega_{p,j}$,
%$\omega_{j}$ and $\gamma_j$ are the strength, position and width of the $j$th vibration,
%respectively.

%
% Figure 2
%
\begin{figure}[t]
\centering
\includegraphics[width=1.0\columnwidth]{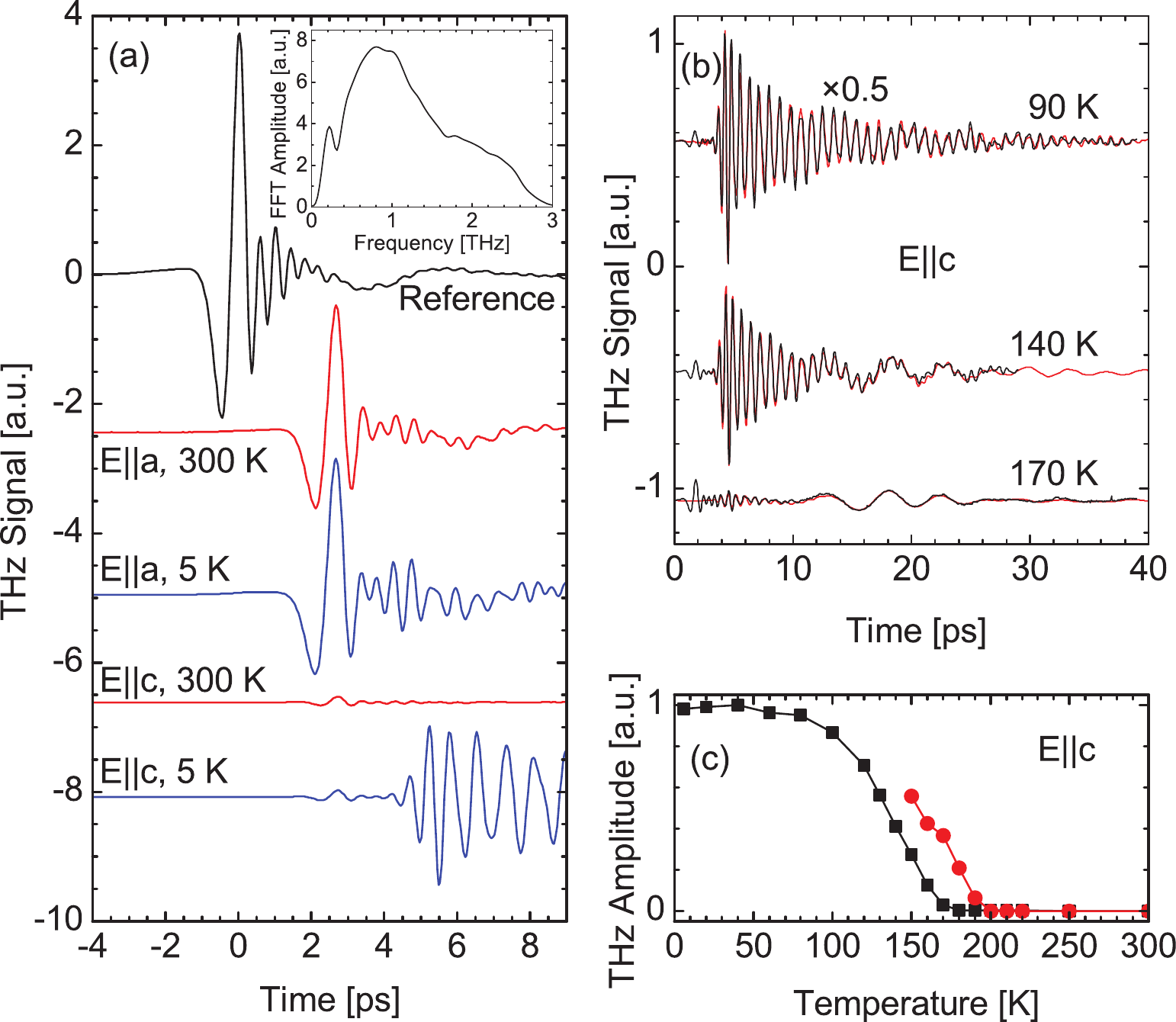}
\caption{Electric field $E(t)$ of THz pulse in the time-domain transmitted through
a 440-$\mu$m thick Sr$_{14}$Cu$_{24}$O$_{41}$ crystal. (a) $E(t)$ polarized along
the {\em a}- and {\em c}-axis at 300 and 5~K.  The reference is recorded without the
crystal. The inset shows the FFT amplitude spectrum of $E(t)$ for the reference.
(b) $E(t)$ (black lines) along the {\em c}-axis shown with the Drude-Lorentz
time-domain fits (red lines) at three different temperatures. (c) Temperature
dependence of the largest peak of the amplitude for the fast oscillation (black
squares) and slow oscillation (red circles) of the THz signal. The slow oscillation
is only shown to 150~K.}%
\label{fig:ef}%
\end{figure}

Figure~\ref{fig:ef}(a) shows the electric field of the THz pulse transmitted through the
Sr$_{14}$Cu$_{24}$O$_{41}$ sample comparing the response for polarizations along the
{\em a}- and {\em c}-axis at high and low temperatures, as well as a reference signal
without a sample.
The Fourier transformed (FFT) amplitude spectrum for the reference is shown in the inset.
For the electric field polarized along the {\em a}-axis, the shape of the THz pulse is
essentially the same at 300 and 5~K, characteristic of insulating behavior along this
direction. However, for the electric field polarized along the {\em c}-axis, the
THz signal is barely present at 300~K, indicative of metallic response at higher temperatures.
At low temperatures a distinct long-lived ringing is observed with a period of $\simeq 1$~ps.
Figure~\ref{fig:ef}(b) shows the transmitted THz pulse polarized along the {\em c}-axis for
three different temperatures with time-domain fits revealing an additional $\simeq 4$~ps oscillation.
The $\simeq 4$~ps period oscillation is observed to appear below $\simeq 200$~K, while the the
$\simeq 1$~ps oscillation below $\simeq 170$~K, as shown in Fig.~\ref{fig:ef}(c).

% At the lowest temperatures the THz oscillation is observed to extend to more than 60~ps, and one notices
% the $\sim$1~ps period oscillation being modulated by a slower $\sim$4~ps period oscillation below the MIT.
% The temperature dependence %the normalized time-domain of the amplitudes of these two oscillations are shown
% in Fig.~\ref{Fig: 1}(e). Above $\sim$80~K, the $\sim$1~ps period signal becomes weaker and vanishes at $\sim$170~K,
% while the $\sim$4~ps period signal remains until $\sim$200~K.

%
% Figure 3 (line weight 4)
%
\begin{figure}[t]
%\centering
\hspace{-0.3cm}
\includegraphics[width=0.95\columnwidth,angle=90]{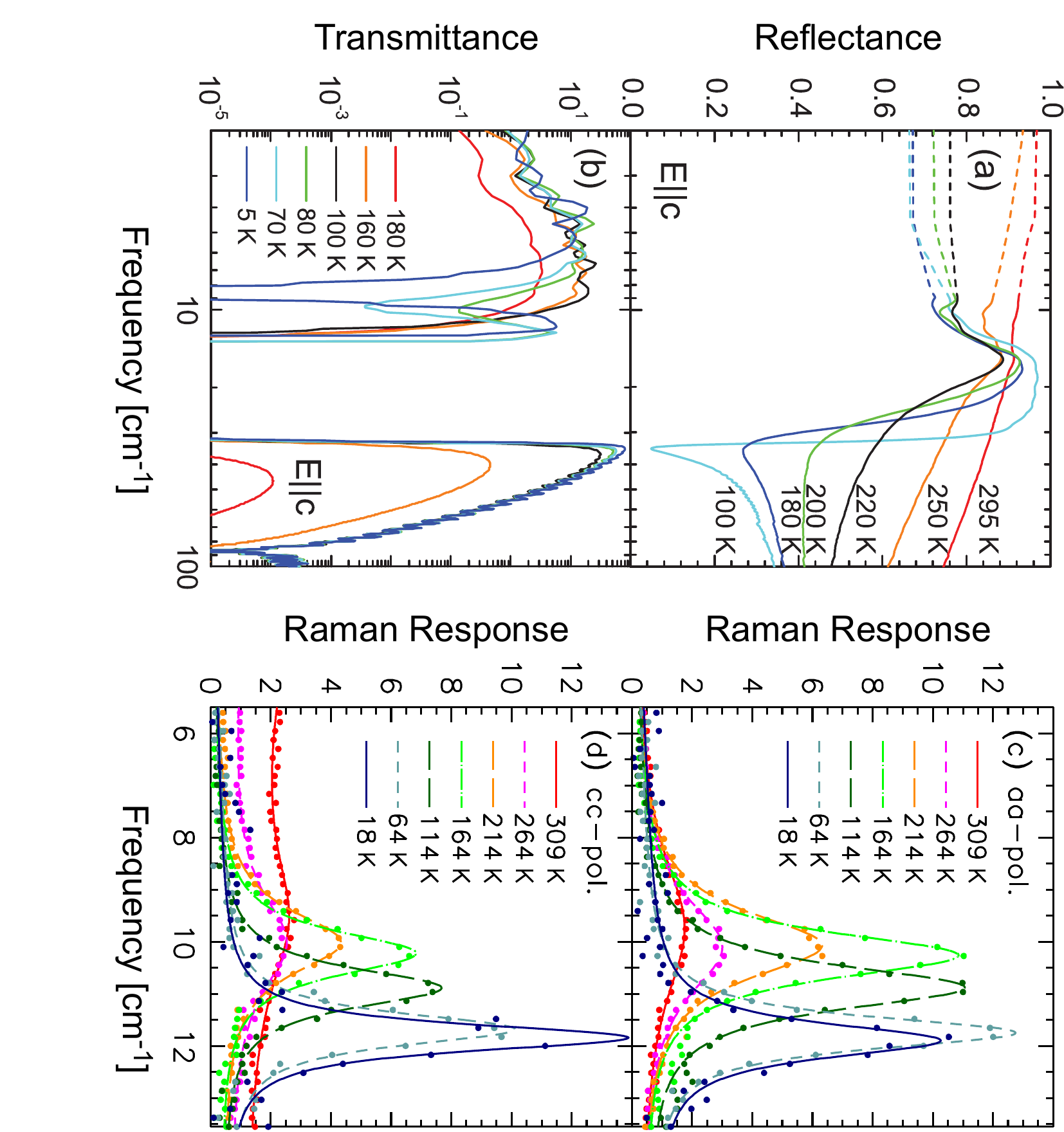}
%\centerline{\includegraphics[width=8.8cm]{figure3.pdf}}
\vspace*{-0.9cm}
\caption{(a) Infrared reflectance of Sr$_{14}$Cu$_{24}$O$_{41}$ for light
polarized along the {\em c}-axis above and below $T_{\rm co}$ where the
dashed lines indicate the $\omega\rightarrow 0$ extrapolations.
(b) Transmittance for light polarized along the {\em c} axis ($T \lesssim T_{\rm co}$).
(c) Raman response in relative units for the {\em aa} polarization, and (d)
{\em cc} polarization at various temperatures.  The Raman data were
adapted from Ref.~\onlinecite{gozar2005a,*gozar2005b}.
}%

\label{fig:raman}%
\end{figure}

The far-infrared reflectance for light polarized along the {\em c}-axis is
shown in Fig.~\ref{fig:raman}(a), while (b) shows the transmittance for the
electric field polarized along the {\em c}-axis of the THz oscillations presented
in Fig.~\ref{fig:ef}; Figs.~\ref{fig:raman}(c) and (d) show the {\em aa} and
{\em cc} Raman responses, respectively \cite{gozar2005a,*gozar2005b}.
A metallic Drude-like behavior is observed in the reflectance above $T_{\rm co}$.
For $T \lesssim T_{\rm co}$ the reflectance changes from a metallic to an
insulating character, allowing a strong vibrational feature to emerge.
%
%This ``filter effect'' becomes even more pronounced in the transmittance data which better
%captures the details of the optical response in the highly transparent regime below
%$\simeq 100$~cm$^{-1}$ at low temperature.
%
For the transmittance shown in Fig.~\ref{fig:raman}(b) there is a sharply
defined window where the transmittance is effectively blocked by at least 5 orders of
magnitude between $\simeq 13 - 33$~cm$^{-1}$ (1~meV $\simeq 8.1$~cm$^{-1}$,
1~THz $\simeq 33.3$~cm$^{-1}$), corresponding to the pronounced slow and
fast oscillations observed in the time-domain (Fig.~\ref{fig:ef}).  As the
temperature is raised, the upper limit of this window becomes less effective
until $T \gtrsim T_{\rm co}$, at which point the entire region becomes
increasingly opaque. In addition, one also notices an absorption at
$\simeq 8$~cm$^{-1}$ which also decreases with increasing temperature.
The loss of the electronic background for $T \lesssim T_{\rm co}$ signals the
transition to a DW ground state and an insulating phase.

%
% Figure 4
%
\begin{figure}[t]
%\centering
\includegraphics[width=0.95\columnwidth]{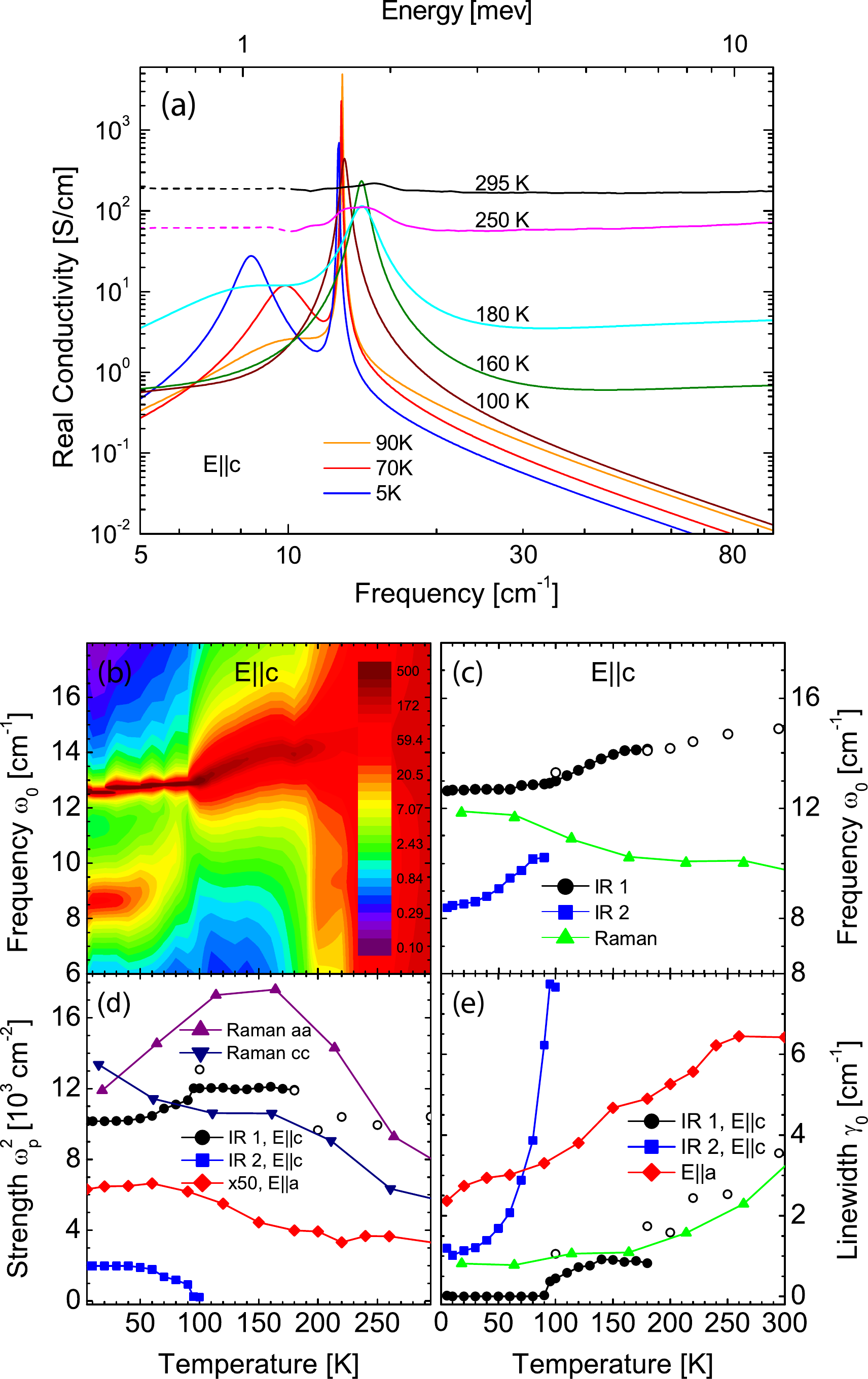}
%\centerline{\includegraphics[width=9.8cm]{figure4.eps}}
\caption{(a) THz-TDS conductivity along the {\em c}-axis of
Sr$_{14}$Cu$_{24}$O$_{41}$ above and below $T_{\rm co}$.  Above 180~K
the conductivity was calculated from the reflectance (see text); the
dashed lines indicate where the conductivity has been determined
from the extrapolations supplied for the reflectance [Fig.~2(a)].
(b) Contour plot of the conductivity along the {\em c}-axis (a) as a function of
frequency and temperature.  (b) Temperature dependence of the two phonon modes observed
in (a) shown together with the Raman-active mode [Figs.~\ref{fig:raman}(c),(d)].
The strengths and linewidths are shown in (d) and (e), respectively, together with that of the
{\em a}-axis phonon mode at $\simeq 57$~cm$^{-1}$. The amplitudes of the Raman data in (d) is
in arbitrary units. The open symbols indicate data obtained from reflectance measurements.}%
\label{fig:SigmaContour}%
\end{figure}

Figure~\ref{fig:SigmaContour}(a) shows the temperature-dependence of the real part
of the optical conductivity along the {\em c}-axis.  A metallic response is observed
for $T \gtrsim T_{\rm co}$, while for $T \lesssim T_{\rm co}$ an insulating response
develops.  The mode at 14.9~cm$^{-1}$ narrows and softens to 12.6~cm$^{-1}$
at low temperature with a slight kink around $T_{\rm co}$ (we refer to this mode as IR1).
The Raman mode seen in {\em aa} and {\em cc} polarizations has the opposite behavior;
its position increases from 9.5 to 12~cm$^{-1}$ upon cooling from $\simeq 300$ to 5~K.
Interestingly, below $T_{\rm co}$, a new infrared mode branches off below 10~cm$^{-1}$
towards lower frequencies, reaching 8.4~cm$^{-1}$ at $\simeq 5$~K; the temperature
dependence of this new mode (which we denote IR2) closely resembles the behavior
reported for the weakly-dispersive  magnetic chain excitations \cite{Matsuda1999}.
Along the {\em a}-axis the optical conductivity is an order of magnitude smaller
than it is along the {\em c} axis, confirming the insulating behavior
perpendicular to the chains and ladders (not shown); in addition to a number
of weak features, a sharp infrared phonon is observed at $\simeq 57$~cm$^{-1}$.
The remaining panels in Figure~\ref{fig:SigmaContour} show the details of the
{\em c}-axis features in the optical conductivity with (b) a contour plot of
optical conductivity as a function of frequency and temperature; (c) the frequency
of IR1 and IR2, as well as the frequency of the {\em aa} and {\em cc} Raman
excitations; their respective (d) strengths and (e) linewidths.  We note that the
total oscillator strength of IR1 and IR2 appears to be conserved
[Fig.~\ref{fig:SigmaContour}(d)].
%

%
% -> I am removing the discussion of this feature because while it is shown it is
% tangential to the main body of work...
% The fitted vibrational parameters of the {\em a}-axis phonon at $\simeq 57$~cm$^{-1}$ are
% also shown in Figs.~\ref{fig:SigmaContour}(d) and (e).  The spectral weight of the
% {\em a}-axis phonon increases in strength for $T < T_{\rm co}$, appearing to track the
% order parameter of the DW transition.
%
% -> I've moved this part to the beginning of the paragraph...
%A low-lying infrared mode, which we will refer to as IR1, is present above and below $T_{\rm co}$ and
%softens with decreasing temperature from 14.9 to 12.6~cm$^{-1}$, with a slight kink around
%$T_{\rm co}$ where the charge order in the ladders and chains sets in.
%

It is tempting to relate these excitations to folded phonon modes due to the $10/7$
superstructure. However, using $c \simeq 27.5$~{\AA} and the sound velocity
$v_s \simeq 13$~km/s \cite{Konig1997}, one would expect these excitations at
energies $\simeq 100$~cm$^{-1}$. This is a factor of ten higher than what we
observe suggesting that a different mechanism is at work.

%Considering the value of the experimentally determined sound velocity
%$v_s \simeq 13$~km/s \cite{Konig1997}, we find that the associated length scale is
%$\xi = v_s / \omega \simeq 1000$~{\AA}.  This value is much larger than the expected
%$c \simeq 27.5$~{\AA} given by the $10/7$ modulation, suggesting that a different
%mechanism is at work.
%
%Girsh correction:
%$v_s \simeq 13$~km/s  í 1.3 10^3 m/s

We then turn our attention to the scenario of the sliding motions of the chain and
ladder subsystems presented in Fig.~\ref{fig:modes}. The energy scale and the relative
frequencies of the Raman and IR1 modes can be understood qualitatively by taking into
account the {\em c}-axis incommensurability between the chain and ladder unit cells,
$c_{\rm ch} / c_{\rm ld} \simeq 0.699$, in accordance with recent x-ray studies
\cite{Gotoh2003,Zimmermann2006} which demonstrated that consideration of the
super space group is mandatory.  We assign $n$, $m$ and $q$ to the corresponding atomic
mass density, unit-cell mass and charge respectively.  Using the available crystallographic
data we obtain without any fitting parameters
\begin{equation}
  \omega_{\rm IR1} = \sqrt{\frac{n_{\rm ch} q_{\rm ch}^2}{\varepsilon_0 \varepsilon_{\infty} m_{\rm ch}}
  \left( 1 + \frac{c_{\rm ld}}{c_{\rm ch}} \frac{m_{\rm ch}}{m_{\rm ld}} \right)}
  \simeq 29.5 \, {\rm cm}^{-1}
  \label{ir1}
\end{equation}
for IR1 \cite{Theodorou1980}. Here the chains and ladders are considered uniformly charged and
$\varepsilon_{\infty} (75\,{\rm cm}^{-1}) \simeq 15$ is the experimentally measured contribution
of all other higher-energy phonons to the dielectric permittivity. All holes were assumed to be
located in the chain system, i.e. $q_{\rm ch} = -1.4\,e$ and $q_{\rm ld} = 2\,e$. This is the ionic complement of  the electronic zone-center plasmons in metals. In general this mode is acoustic with
a diffusive character at long wavelengths but it becomes gapped due to restoring Coulomb forces
if the IC systems are oppositely charged \cite{Theodorou1978,Theodorou1980,Axe1982,Finger1983}.
The Raman mode corresponds to the out-of-phase oscillation of the chain layers, phase
shifted by $\pi$ along the {\em b}-axis, with the ladders at rest (Fig.~\ref{fig:modes}).
Its frequency can be estimated by Eq.~(\ref{ir1}) in the $m_{\rm ld} \rightarrow \infty$ limit.
Hence, $\omega_{\rm R} \simeq 0.85\,\omega_{\rm IR1}$ which is in good agreement with the
experimental observations.
Removal of free carriers with decreasing temperature due to the activated nature
of the conductivity would reduce screening effects, leading to the Raman mode hardening at
low temperatures, also in agreement with our observations.

% Girsh comment: Eq.~(\ref{ir1}) represents the energy of the {\em c}-axis
% [sliding mode associated with the]
% out-of-phase oscillation of the chain and ladder planes
% and is the ionic complement of the electronic zone-center plasmons in metals.
%
% The Raman mode is identified as the sliding motion of the chain planes, with
% phases shifted by $\pi$ along the {\em b}-axis with the ladders at rest.  This
% mode carries no dipole moment and corresponds to $\omega_{\rm R} = \omega_{\rm IR 1}
% (m_{\rm ld} \rightarrow \infty) = 0.85\,\omega_{\rm IR 1}$ in good agreement with the
% experimental observations. See Fig.~\ref{fig:ModeSketches} for an illustration of the IR and Raman modes.
%
% Girsh comment: The Raman mode is identified as the out-of-phase oscillation of the chain planes,
% with phases shifted by $\pi$ along the {\em b} axis with the ladders at rest.
%This mode carries no dipole moment and its frequency can be estimated by eq. (2) in the
% $m_{\rm ld} \rightarrow \infty$ limit. Hence, $\omega_{\rm R} \simeq 0.85\,\omega_{\rm IR 1}$
% in good agreement with the experimental observations.

We suggest that the origin and energy of the IR2 mode can be understood by
considering the effects of quasi-2D charge ordering in the chains \cite{Matsuda1999}. Once
the long-range hole ordering in the chains sets in below $\simeq 100$~K restoring Coulomb
forces will oppose the out-of-phase oscillation of adjacent chains.  Above 100~K this
excitation is expected to have a vanishingly small energy because of the short range
charge correlations along the {\em a}-axis resulting in the absence of net restoring
electrostatic forces in the disordered state.
For a long-range sinusoidal charge modulation, the energy of this mode is
\begin{equation}
   \omega_{\rm IR2}^2 \propto \sqrt{\frac{\lambda}{\pi d}} \,\exp\left( {-\frac{\pi d}{\lambda}} \right)
   \frac{\delta q^2}{m_{\rm ch} c_{\rm ch} \lambda^2},
   \label{ir2}
\end{equation}
with a proportionality  factor of the order of unity.  Here $d$, $\lambda$ and $\delta q$ are
the distance between two chains, the wavelength and depth of the harmonic charge modulation,
respectively.  In Sr$_{14}$Cu$_{24}$O$_{41}$ this charge modulation is not a simple sinusoid;
however, to first order a harmonic approximation may be used for two coupled chains.
Taking $\lambda \simeq 5\, c_{\rm ch}$, $d \simeq 5.7$~{\AA} and an average charge
modulation depth $\langle \delta q \rangle \simeq 0.37\,{e}$, we find
$\omega_{\rm IR2} \approx 3$~cm$^{-1}$, again consistent with the experimental data. From
Eq.~(\ref{ir2}) it is seen that IR2 is a direct probe of the charge order.  The situation is
quite similar to the undoped La$_{6}$Ca$_{8}$Cu$_{24}$O$_{41}$ where the staggered chain
arrangement along the {\em a}-axis should generate restoring Coulomb forces $\pi$-shifted
oscillations of adjacent chains and the analog of the IR2 mode in Sr$_{14}$Cu$_{24}$O$_{41}$
is expected to be present at all temperatures.  This is in agreement with our experimental
observations.  A quantitative analysis of the IR2 mode and comparison to crystals of the
``14-24-41'' family will be the topic of a future study.

In conclusion, we propose that the low-energy Raman and IR1 excitations in Sr$_{14}$Cu$_{24}$O$_{41}$
originate from sliding motions of the IC chains and ladders which are gapped by Coulomb
interactions due to the net charge carried by these sub-systems.  Long-range charge ordering in the chains
will further generate low-energy infrared activity and we suggest this to be at the origin of the new
IR2 mode observed in the time-domain THz data below $\simeq 100$~K.  The energy of this mode is as such a direct
probe of the charge modulation depth as well as of the quasi-2D hole ordering pattern in the chain subsystem.

We gratefully acknowledge useful discussions with T.~Maurice Rice and Jason Hancock.
We would like to thank H.~Eisaki for providing us with samples. Work at Brookhaven was
supported by the U.S. Department of Energy, Office of Basic Energy Sciences, Division
of Materials Sciences and Engineering under Contract No. DE-AC02-98CH10886. Work at
Rutgers was supported by NSF DMR-1104884. Thanks to Prof.~H.~L.~Bhat of IISc, Bangalore for
the Indo-French collaborative project, CEFIPRA under project No.3408-4,  which supported
the crystal growth work at CEA Grenoble.

%%%%%%%%%%%%%%%%%%%%%%%%%%%%%%%%%%%%%%%%%%%%%%%%%%%%%%%%%%%%%%%%%%%%%%%%%%%%%%
%
% References
%
%\bibliography{ladders}

\providecommand{\noopsort}[1]{}\providecommand{\singleletter}[1]{#1}

\end{document}